\documentclass[%
 aip,
 jmp,%
 amsmath,amssymb,
preprint,%
]{revtex4-1}

\usepackage{graphicx}
\usepackage{dcolumn}
\usepackage{bm}

\begin{document}


\title{Time-dependent restricted-active-space self-consistent-field singles method for many-electron dynamics}

\author{Haruhide Miyagi}
\author{Lars Bojer Madsen}
\affiliation{Department of Physics and Astronomy, Aarhus University, 8000 Aarhus C, Denmark}

\date{\today}

\begin{abstract}
The time-dependent restricted-active-space self-consistent-field singles (TD-RASSCF-S) method is presented for investigating TD many-electron dynamics in atoms and molecules. Adopting the SCF notion from the muticonfigurational TD Hartree-Fock (MCTDHF) method and the RAS scheme (single-orbital excitation concept) from the TD configuration-interaction singles (TDCIS) method, the TD-RASSCF-S method can be regarded as a hybrid of them. We prove that, for closed-shell $N_{\rm e}$-electron systems, the TD-RASSCF-S wave function can be fully converged using only $N_{\rm e}/2+1\le M\le N_{\rm e}$ spatial orbitals. Importantly, based on the TD variational principle, the converged TD-RASSCF-S wave function with $M= N_{\rm e}$ is more accurate than the TDCIS wave function. The accuracy of the TD-RASSCF-S approach over the TDCIS is illustrated by the calculation of high-order harmonic generation spectra for one-dimensional models of atomic helium, beryllium, and carbon in an intense laser pulse. The electronic dynamics during the process is investigated by analyzing the behavior of electron density and orbitals. The TD-RASSCF-S method is accurate, numerically tractable, and applicable for large systems beyond the capability of the MCTDHF method.
\end{abstract}

\maketitle

\section{Introduction}

The time-dependent (TD) many-electron problem involving non-perturbative interactions and including one or more continua remains a tremendous challenge for theory. The current and future developments of intense femtosecond and ultrashort attosecond laser pulses~\cite{Krausz2009,pascal2012} as well as pulsed electron beams~\cite{zewail2008,zewail2010,miller2011} require formulation of reliable explicitly TD \textit{ab initio} theories to resolve the electron correlation encoded in experimental results, and to elucidate electron dynamics on their natural length and timescales. The description of a theory with properties along these lines is the purpose of this work.

Consider an $N_{\rm e}$-electron system governed by a TD Hamiltonian, $H(t)$. Based on the spin restricted ansatz, the wave function is constructed from $N$ electronic configurations using $M$ spatial orbitals. Computations of many-electron dynamics induced by strong laser pulses or collision processes require large simulation volumes and many  basis functions, $N_{\rm b}$ (in the one-dimensional (1D) 
model calculations considered below $N_{\rm b}  =O(10^3)$), which in general makes both $M$ and $N$ very large. To reduce the computational cost, the muticonfigurational TD Hartree-Fock (MCTDHF) method \cite{Meyer1990,Beck2000,Caillat2005,Meyer2010} is based on the self-consistent-field (SCF) scheme, by which accurate wave functions can be obtained with a relatively small number of orbitals, $M=O(N_{\rm e})$. Due to the full-CI expansion, however, as $N_{\rm e}$ increases, the computation becomes difficult with exponential increase in $N=O(M^{N_{\rm e}})$. On the other hand, the TD configuration-interaction singles (TDCIS) method~\cite{Rohringer2006,Gordon2006,Rohringer2009,Greenman2010,Pabst2012,Pabst2012b,Pabst2013} simply uses the time-independent HF occupied and virtual orbitals. Although the number of orbitals is huge, $M=O(N_{\rm b})$, the number of electronic configurations depends linearly on $N_{\rm e}$, $N=O(N_{\rm b}N_{\rm e})$. The TDCIS method is thus numerically tractable for large systems beyond the reach of the MCTDHF method, and thereby currently succeeds in analyzing attosecond light absorption~\cite{Pabst2012b} and high-order harmonic generation (HHG) processes~\cite{Pabst2013}.

To extend the applicability of SCF based methods, we recently introduced the TD restricted-active-space SCF (TD-RASSCF) method~\cite{Miyagi2013,Miyagi2014a}. The unfavorable scaling of the MCTDHF method with $N$ is cured by the RAS scheme \cite{Helgaker2000,Hochstuhl2012}, i.e., by taking into account only important  configurations. The aim of this work is to focus on the TD-RASSCF singles (-S) method. In a sense, this method is an extension of the TDCIS method by incorporating the SCF scheme. Due to the hybrid property, the TD-RASSCF-S wave function is expected to be accurate with a small number of orbitals $M=O(N_{\rm e})$ and configurations $N=O(N_{\rm e}^2)$. In general, however, adding more orbitals makes the wave function more accurate and the computation more expensive. It is hence {\it a priori} unclear how many orbitals are needed to make the wave function sufficiently converged and more accurate than the TDCIS wave function. In Sec.~\ref{Theory}, we answer these questions by proving a theorem which states that, for closed-shell systems, the TD-RASSCF-S wave function can be fully converged with only $N_{\rm e}/2+1\le M\le N_{\rm e}$ orbitals. We show that the converged TD-RASSCF-S wave function with $M=N_{\rm e}$ is more accurate than the TDCIS wave function. These properties make the TD-RASSCF-S method very attractive for applications to non-perturbative TD many-electron dynamics. We know of no other TD theory where the question about the number of orbitals needed for convergence at a given level of approximation can be answered. In Sec.~\ref{Numerical experiments and discussion}, by carrying out numerical experiments, we then demonstrate the accuracy of the TD-RASSCF-S method. By analyzing the TD electron density and the behavior of orbitals during the nonperturbative high-order harmonic generation (HHG) process, we consider how the TD-RASSCF-S method takes into account the electron correlation accurately.

\begin{figure*}
\begin{center}
\begin{tabular}{c}
\resizebox{160mm}{!}{\includegraphics{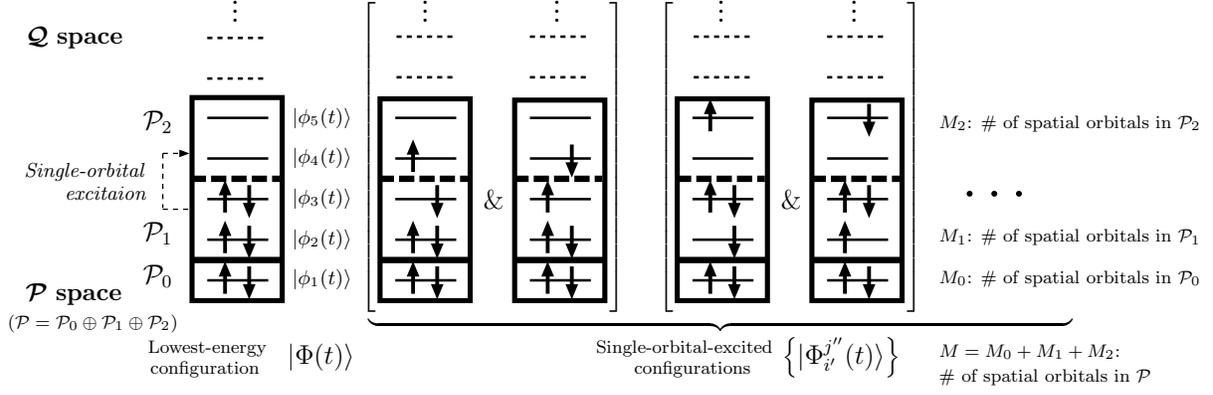}}
\end{tabular}
\caption{
\label{fig_directsum-S}
Illustration of key concepts in the TD-RASSCF-S method. The wave function is composed of the spin orbitals $|\phi_i(t)\rangle\otimes|\sigma\rangle$ ($i=1,\cdots,M$, and $\sigma=\uparrow,\downarrow$). The spatial orbitals, $\big\{|\phi_i(t)\rangle\big\}_{i=1}^M$, are numbered in ascending order from the lowest energy. The $\mathcal{P}$ space spanned by the spatial orbitals consists of three subspaces: an inactive-core space, $\mathcal{P}_0$, and two active spaces, $\mathcal{P}_1$ and $\mathcal{P}_2$, between which single-orbital transitions are allowed. The rest of the single-particle Hilbert space spanned by the virtual orbitals is referred to as $\mathcal{Q}$ space. The number of spatial orbitals in the $\mathcal{P}_0$, $\mathcal{P}_1$, and $\mathcal{P}_2$ spaces are expressed by $M_0$, $M_1$ and $M_2$, respectively, and the total number by $M=M_0+M_1+M_2$. The illustration shows a six-electron system ($N_{\rm e}=6$) with $(M_0,M_1,M_2)=(1,2,2)$, and $M=5$.
}
\end{center}
\end{figure*}

\section{\label{Theory} Theory}

We start by introducing essential concepts for defining the TD-RASSCF-S method. The wave function consists of the TD spin orbitals, $|\phi_i(t)\rangle\otimes|\sigma\rangle$ ($i=1,\cdots,M$, and $\sigma=\uparrow,\downarrow$ denoting the spin states), with a set of $M(\ge N_{\rm e}/2)$ spatial orbitals, $\big\{|\phi_i(t)\rangle\big\}_{i=1}^M$. Let $\mathcal{P}$ be the space spanned by the spatial orbitals and $\mathcal{Q}$ the rest of the single-particle Hilbert space. As illustrated in Fig.~\ref{fig_directsum-S}, $\mathcal{P}$ is divided into three subspaces: inactive-core space, $\mathcal{P}_0$, and two active spaces, $\mathcal{P}_1$ and $\mathcal{P}_2$, between which single-orbital excitations are allowed. Let $M_0$, $M_1$, and $M_2$ denote the numbers of spatial orbitals in $\mathcal{P}_0$, $\mathcal{P}_1$, and $\mathcal{P}_2$, respectively (hence, $M=M_0+M_1+M_2$). For simplicity, we suppose a closed-shell system (so that $N_{\rm e}$ is even) and also the condition $M_0+M_1=N_{\rm e}/2$. The TD-RASSCF-S wave function is expanded in terms of normalized Slater determinants composed of $N_{\rm e}$ TD spin orbitals,
\begin{eqnarray}
|\Psi(t)\rangle
=
C_0(t)|\Phi(t)\rangle
+
\sum_{i'j''}C_{i'}^{j''}(t)|\Phi_{i'}^{j''}(t)\rangle,
\label{expression1}
\end{eqnarray}
where orbitals denoted with single (double) primed index $i'$ ($j''$) belong to $\mathcal{P}_1$ ($\mathcal{P}_2$). The lowest-energy configuration is represented by $|\Phi(t)\rangle$, from which single-orbital-excited configurations are obtained. We define 
\begin{eqnarray}
|\Phi_{i'}^{j''}(t)\rangle\equiv \big(c_{j''\uparrow}^{\dagger}c_{i'\uparrow}+c_{j''\downarrow}^{\dagger}c_{i'\downarrow}\big)|\Phi(t)\rangle,
\end{eqnarray}
where $c_{i\sigma}$ ($c^{\dagger}_{i\sigma}$) is the annihilation (creation) operator of an electron in the spin orbital $|\phi_i(t)\rangle\otimes|\sigma\rangle$. By numbering the spatial orbitals in ascending order from the lowest energy as shown in Fig.~\ref{fig_directsum-S}, the summations in Eq.~\eqref{expression1} are taken for $i'\equiv M_0+i$ ($i=1,\cdots,M_1$) and $j''\equiv M_0+M_1+j$ ($j=1,\cdots,M_2$). 

To compute ground-state wave functions, Refs.~\onlinecite{Dacre1975,Malykhanov1985} presented a time-independent MCSCF method based on the same expansion style as Eq.~\eqref{expression1}. Our main purpose is, however, the time propagation of the wave function, for which we originally derived the equations of motion for the CI-expansion coefficients and orbitals. The Dirac-Frenkel-McLachlan TD variational principle \cite{Dirac1930, Frenkel1934, McLachlan1964, Lubich2008} gives a prescription (The details are given elsewhere~\cite{Miyagi2013,Miyagi2014a}). The TD variational principle provides the best approximation within a given set of variational parameters and gives a more accurate wave function by adding more parameters. Hence we need to use as many orbitals as possible to compute the observables of interest within a tolerance of convergence. Exceptionally, however, the TD-RASSCF-S wave function is converged for $N_{\rm e}/2+1\le M\le N_{\rm e}$, which is stated as a theorem and can be proven as follows:

\vspace{3mm}
\noindent{\bf Theorem}
\vspace{1mm}

{For closed-shell systems ($N_{\rm e}$ is even), the TD-RASSCF-S method satisfying $M_0+M_1=N_{\rm e}/2$ and $M_1\le M_2$ gives a wave function which is invariant with respect to the value of $M_2$.} \\

\vspace{3mm}
\noindent{\bf Proof}
\vspace{1mm}

Consider orbital rotations in the $\mathcal{P}_1$ and $\mathcal{P}_2$ spaces separately: (for brevity, explicit time-dependence is dropped in our notation)
\begin{eqnarray}
c_{i'\sigma} &\to& \sum_{j'}u_{i'j'}c_{j'\sigma}, \\
c_{i''\sigma} &\to& \sum_{j''}v_{i''j''}c_{j''\sigma}.
\end{eqnarray}
Let ${\bf U}$ be the $M_1\times M_1$ unitary matrix with $({\bf U})_{ij}=u_{i'j'}$, $i'= M_0+i$, and $j'= M_0+j$ ($i,j=1,\cdots,M_1$), and let ${\bf V}$ denote the $M_2\times M_2$ unitary matrix with $({\bf V})_{ij}=v_{i''j''}$, $i''= M_0+M_1+i$, and $j''= M_0+M_1+j$ ($i,j=1,\cdots,M_2$). Similarly, let ${\bf C}$ be the $M_2\times M_1$ matrix with $({\bf C})_{ji}=C_{i'}^{j''}$, and let ${\bm \Phi}$ denote the $M_1\times M_2$ matrix with $({\bm \Phi})_{ij}=|\Phi_{i'}^{j''}\rangle$ ($i=1,\cdots,M_1$, and $j=1,\cdots,M_2$). The effect of the orbital rotations can then be expressed in matrix form: ${\bm \Phi}\to {\bf U}{\bm \Phi}{\bf V}^{\dagger}$.

The single-orbital-excited configurations in Eq.~\eqref{expression1} are transformed as follows:
\begin{eqnarray}
\sum_{{i'}{j''}}C_{i'}^{j''}
|\Phi_{i'}^{j''}\rangle
&=&
\sum_{ij}
({\bf C})_{ji}({\bf \Phi})_{ij}
\nonumber \\
&=&
\sum_{ij}
({\bf C})_{ji}({\bf U}^{\dagger}{\bf U}{\bf \Phi}{\bf V}^{\dagger}{\bf V})_{ij}
\nonumber \\
&=&
\sum_{ij}
\big({\bf V}{\bf C}{\bf U}^{\dagger}\big)_{ji}
({\bf U}{\bf \Phi}{\bf V}^{\dagger})_{ij}.
\label{expression3}
\end{eqnarray} 

Since ${\rm rank}\:{\bf C}=\min\big\{M_1,M_2\big\}$, it has no effect on the wave function to reduce the dimension of the $\mathcal{P}_2$ space as long as $M_1<M_2$. The wave function is therefore invariant with respect to the value of $M_2(\ge M_1)$. $\blacksquare$

\vspace{3mm}

Equation \eqref{expression3} expresses a property of SCF theories called parametric redundancy \cite{Helgaker2000,Lubich2008}, i.e., a set of orbital rotations, ${\bm \Phi}\to {\bf U}{\bm \Phi}{\bf V}^{\dagger}$, accompanied by the proper transformation of the CI-expansion coefficients, ${\bf C} \to {\bf V}{\bf C}{\bf U}^{\dagger}$, leaves the wave function invariant. 

The time evolution of the orbital rotations, ${\bf U}(t)$ and ${\bf V}(t)$, is uniquely determined by solving the equations of motion for two different pictures defined at $t=0$ by choosing arbitrary ${\bf U}(0)$ and ${\bf V}(0)$. Thus Eq.~\eqref{expression3} is correct at any time. Also note that, if $M_0+M_1>N_{\rm e}/2$, the CI-expansion coefficients can not be labeled by two indices like $C_{i'}^{j''}$ in Eq.~\eqref{expression1}, which results in the break-down of the theorem because the proof is based on representing the coefficients in matrix form. The single-orbital excitation ansatz is also essential for the matrix form, which is again broken in multi-orbital excitation schemes.

We now come back to the relation to the TDCIS method and the consequence of the theorem. Let $N_{\rm b}$ denote the number of basis functions to expand each orbital. First consider the case: $(M_0,M_1,M_2)=(0,N_{\rm e}/2,N_{\rm b}-N_{\rm e}/2)$, i.e., $\mathcal{P}_1\oplus \mathcal{P}_2$ covers the entire single-particle Hilbert space and there is no $\mathcal{Q}$ space. The TD-RASSCF-S and the TDCIS wave functions are then composed of the same number and kind of electronic configurations, but the TD-RASSCF-S wave function is more accurate because of the variational optimization of the orbitals. By the theorem, the dimension of $\mathcal{P}_2$ can be reduced to $M_2=N_{\rm e}/2$, leaving the TD-RASSCF-S wave function invariant. Hence, using the condition $(M_0,M_1,M_2)=(0,N_{\rm e}/2,N_{\rm e}/2)$, the TD-RASSCF-S method is numerically tractable, and the wave function is more accurate than the TDCIS wave function. Here the wording `accurate' is based on the TD variational principle and defined such that adding more variational parameters results in more `accurate' wave function.

\begin{figure*}[!]
\begin{center}
\begin{tabular}{c}
\resizebox{140mm}{!}{\includegraphics{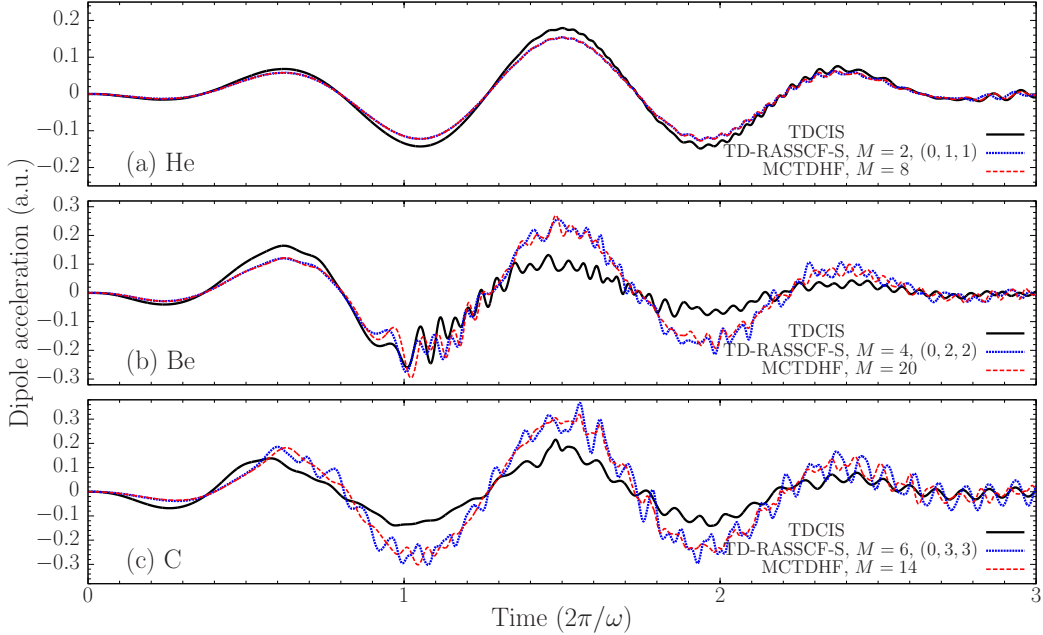}}
\end{tabular}
\caption{
\label{fig_Be_dipole}
Laser-induced dipole acceleration $\langle\Psi(t)|D|\Psi(t)\rangle$ (see text) of the 1D model atoms: (a) helium ($N_{\rm e}=2$), (b) beryllium ($N_{\rm e}=4$), and (c) carbon ($N_{\rm e}=6$). Each panel includes the list of methods, if necessary with the number of spatial orbitals, $M$, and the partitioning, $(M_0,M_1,M_2)$. In the TDCIS calculation, all possible single-orbital excitations from the occupied HF orbitals to the virtual orbitals were taken into account. The laser pulse used in the computation is specified by Eq.~\eqref{lalser}. 
}
\end{center}
\end{figure*}

\begin{figure*}[!]
\begin{center}
\begin{tabular}{c}
\resizebox{140mm}{!}{\includegraphics{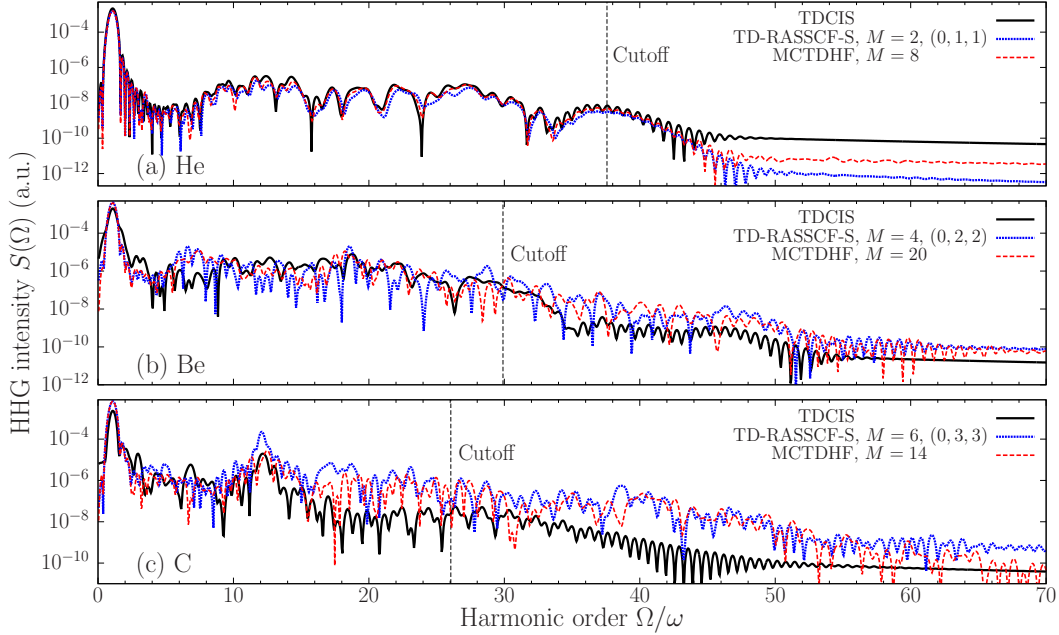}}
\end{tabular}
\caption{
\label{fig_Be_HHG}
HHG spectra of the 1D model atoms: (a) helium ($N_{\rm e}=2$), (b) beryllium ($N_{\rm e}=4$), and (c) carbon ($N_{\rm e}=6$). The panels (a), (b), and, (c), respectively, correspond to those in Fig.~\ref{fig_Be_dipole}. The cutoff energies in the HHG spectra are estimated to be $37.6\omega$, $29.9\omega$, and $26.0\omega$ for the helium, beryllium, and carbon atoms, respectively, as shown by the vertical dotted lines (see text). 
}
\end{center}
\end{figure*}

\begin{figure*}[!]
\begin{center}
\begin{tabular}{c}
\resizebox{140mm}{!}{\includegraphics{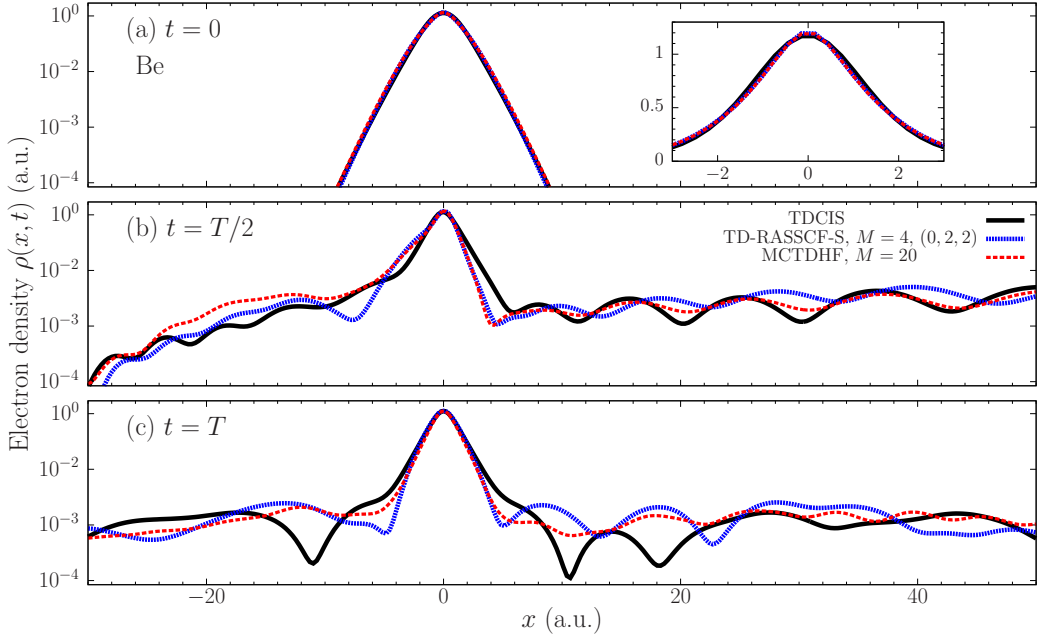}}
\end{tabular}
\caption{
\label{fig_HeBeC_density}
Snapshots of the electron density $\rho(x,t)$ in the 1D beryllium atom at $t=0$ (a), $T/2$ (b), and $T$ (c), computed by the TDCIS, TD-RASSCF-S, and MCTDHF methods. The inset in (a) displays a magnification around the nucleus in linear scale.
}
\end{center}
\end{figure*}

\begin{figure*}[!]
\begin{center}
\begin{tabular}{c}
\resizebox{140mm}{!}{\includegraphics{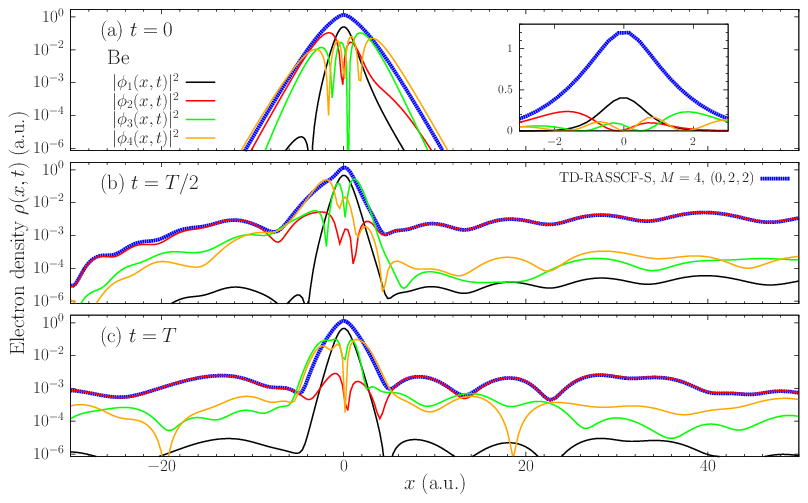}}
\end{tabular}
\caption{
\label{fig_Be_density}
Snapshots of the electron density $\rho(x,t)$ (dotted (blue) lines) in the 1D beryllium atom at $t=0$ (a), $T/2$ (b), and $T$ (c) computed by the TD-RASSCF-S method with $(M_0,M_1,M_2)=(0,2,2)$ (the same as the doted blue lines in Fig.~\ref{fig_HeBeC_density}). Each panel also includes solid thin lines representing the norm squared of spatial orbitals $|\phi_i(x,t)|^2$ ($i=1,2,3,$ and $4$). The spatial orbitals are numbered in ascending order from the lowest energy and $i=1,2$ ($i=3,4$) belong to $\mathcal{P}_1$-space ($\mathcal{P}_2$-space). The inset in (a) displays a magnification around the nucleus in linear scale.
}
\end{center}
\end{figure*}

\section{\label{Numerical experiments and discussion} Numerical experiments and discussion}

To demonstrate the performance of the TD-RASSCF-S method, we carried out test calculations for 1D model atoms. In a $[-300,300(\equiv L)]$ simulation box, the system is described by the TD Hamiltonian: The one-body operator is (atomic units are used throughout)
\begin{eqnarray}
h(x,t)=-\frac{1}{2}\frac{d^2}{dx^2}+V(x)+xF(t)-iW(x),
\label{one-body}
\end{eqnarray}
where $V(x)=-Z/\sqrt{x^2+1}$ with $Z=N_{\rm e}=2,4,$ and $6$ for mimicking atomic helium \cite{Hochstuhl2010,Balzer2010,Balzer2010b}, beryllium \cite{Miyagi2013,Hochstuhl2010b}, and carbon, respectively. Although the lowest-energy state formed by six electrons in the 1D potential well is closed-shell which differs from the actual open-shell 3D carbon, it will be natural and systematic to name the system 1D carbon. The complex absorbing potential function~\cite{Kosloff1986cap} is defined by $W(x)=1-\cos\Big\{\pi(|x|-x_{\rm cap})/\big[2(L-x_{\rm cap})\big]\Big\}$ with $x_{\rm cap}=250$ for $|x|>x_{\rm cap}$ and zero otherwise. Within the framework of the dipole approximation, the laser-electron interaction is represented in the length gauge, $xF(t)$, but the SCF scheme ensures the gauge invariance, i.e., the use of the velocity or acceleration gauge gives no change to the dynamics~\cite{Miyagi2014a,Sato2013,Gauge}. On the other hand, the two-body operator representing the electron-electron repulsion is defined as
\begin{eqnarray}
v(x_1,x_2)=\frac{1}{\sqrt{(x_1-x_2)^2+1}}.
\end{eqnarray}
The TDCIS, TD-RASSCF-S, and MCTDHF calculations were carried out using discrete-variable-representation \cite{Light1985} with $N_{\rm DVR}=2048$ quadrature points associated with Fourier basis functions. Each calculation commenced with imaginary-time relaxation \cite{Kosloff1986} to obtain the ground-state wave function which is the HF state in the TDCIS method. The MCTDHF method gives the most accurate ground-state wave function, and the TD-RASSCF-S method the second most accurate (see Ref.~\onlinecite{Miyagi2014a} which includes details about the ground-state energies obtained from the HF, TD-RASSCF-S, and MCTDHF methods). The calculation then proceeded with real-time propagation under the laser pulse, $F(t)\equiv -dA(t)/dt$, with the vector potential
\begin{eqnarray}
A(t)=\frac{F_0}{\omega}\sin^2\left(\frac{\pi t}{T}\right)\sin\omega t,
\hspace{5mm} (0\leq t\leq T),
\label{lalser}
\end{eqnarray}
and the electric field strength, $F_0=0.0755$ ($2.0\times10^{14}$ Wcm$^{-2}$), the angular frequency, $\omega=0.0570$ ($800$ nm), and the pulse duration, $T=331$ ($3$ cycles). More details of the calculation are given elsewhere~\cite{Miyagi2013,Miyagi2014a}.

To investigate the laser-induced dynamics, Fig.~\ref{fig_Be_dipole} displays the dipole accelerations $\langle\Psi(t)|D|\Psi(t)\rangle$, where $D= -\sum_{\kappa=1}^{N_{\rm e}}dV(x_{\kappa})/dx_{\kappa}$. The figure shows that the TD-RASSCF-S results agree better with the MCTDHF references in both helium, beryllium, and carbon. Figure~\ref{fig_Be_HHG} gives the corresponding HHG spectra computed as the norm squared of the Fourier transformation of the dipole acceleration~\cite{Baggesen2011}. For the helium, beryllium and carbon atoms, the MCTDHF calculations were carried out with $M=8$, $20$, and $14$ spatial orbitals, respectively. Accurate convergence was checked for the helium atom by comparing the result to the direct solution to the TD Schr{\"o}dinger equation (TDSE). For the beryllium and carbon atoms, direct solution of the TDSE is impossible, so the MCTDHF results could not be compared with exact results and may require more orbitals for convergence. On the other hand, the TD-RASSCF-S calculations used the partitioning, $(M_0,M_1,M_2)=(0,N_{\rm e}/2,N_{\rm e}/2)$, for which exact convergence is ensured by the theorem. Based on the classical model \cite{Krause1992,Schafer1993,Corkum1993} for HHG (see also quantum mode in Ref.~\onlinecite{Lewenstein1994}), the cutoff energies in the HHG spectra are estimated to be $3.17U_{\rm p}+I_{\rm p}=37.6\omega$, $29.9\omega$, and $26.0\omega$ for the helium, beryllium, and carbon atoms, respectively, and indicated by vertical dotted lines in Fig.~\ref{fig_Be_HHG}. Here $U_{\rm p}=F_0^2/(4\omega^2)=0.439$ is the ponderomotive energy (time-averaged energy of a free electron quivering in the laser field). The first ionization potentials $I_{\rm p}=0.750$, $0.313$, and $0.093$ for the helium, beryllium, and carbon atoms, respectively, are estimated based on Koopmans' theorem \cite{Helgaker2000}.

First look at the dipole accelerations in Fig.~\ref{fig_Be_dipole}. For every atom, while the TD-RASSCF-S method reasonably reproduces the MCTDHF results, the TDCIS method gives obvious deviations. Accordingly, the HHG spectra in Fig.~\ref{fig_Be_HHG} given by the TD-RASSCF-S and MCTDHF calculations are in good agreement over the whole region, while the TDCIS method clearly underestimates the HHG intensity above and even around the cutoff. Because of the lack of multi-orbital excitations, the failure of the TDCIS method tends to be pronounced for larger systems showing unclearer cutoff due to smaller ionization potential and larger polarizability. The TD-RASSCF-S method likewise includes only the single-orbital excitations but, owing to the orbital optimization, succeeds in reproducing the MCTDHF result. Recall that, based on the TD variational principle, the MCTDHF result is the most accurate, which is followed by the TD-RASSCF-S and TDCIS results, in this order. The numerical experiment verifies this fact. Also note that the gauge independence is another striking superiority of the TD-RASSCF-S method to the TDCIS method which is gauge dependent~\cite{Miyagi2014a}. Without concerns about the convergence with respect to $M$, the TD-RASSCF-S method therefore gives reasonably accurate and gauge independent results for large systems with practical computational costs. 

To more directly analyze the laser-induced dynamics, Fig.~\ref{fig_HeBeC_density} displays the electron density $\rho(x,t)$ in the 1D beryllium atom. During and after the interaction with the laser pulse, $t=T/2$ and $T$, the electron densities obtained from the MCTDHF and TD-RASSCF-S computations are particularly in good agreement beside the nucleus, around $1<|x|<5$, where the TDCIS results clearly differ from them. The carbon results (not displayed) show a similar trend. The accurate description of the electron density in the vicinity of the nucleus explains why the TD-RASSCF-S gives accurate results for the HHG spectrum (Fig.~\ref{fig_Be_dipole}). Some important details of electronic structure in the atoms may be well described in the TD-RASSCF-S calculation but are missing in the TDCIS approach.

Finally, we look into the behavior of each orbital to understand the working mechanism of the TD-RASSCF-S method. Figure~\ref{fig_Be_density} shows the electron density with the norm squared of spatial orbitals $|\phi_i(x,t)|^2$ ($i=1,2,3,$ and $4$) in the TD-RASSCF-S calculation, where orbitals with $i=1,2$ ($i=3,4$) belong to $\mathcal{P}_1$ space ($\mathcal{P}_2$ space).  Although the orbitals are not unique because they can be unitary transformed within each subspace, it is clear from the figure that the $\mathcal{P}_1$-space orbitals are mainly responsible for describing the localized core around the nucleus and unlocalized ejected electrons far away from the nucleus. Around the nucleus, on the other hand, the $\mathcal{P}_2$-space orbitals have important contributions to the wave function. By explicitly including the single-orbital excitations from $\mathcal{P}_1$ to $\mathcal{P}_2$, the TD-RASSCF-S method takes into account the electron correlation around the nucleus and succeeds in the accurate time propagation.

\section{Summary}

In this work, we have presented the TD-RASSCF-S method as an extension or a hybrid of the TDCIS and MCTDHF methods. For closed-shell systems, the TD-RASSCF-S method shows a special convergence property: the wave function is converged for $N_{\rm e}/2+1\le M\le N_{\rm e}$. By proving and exploiting it, we showed that the converged wave function with $M= N_{\rm e}$ is more accurate than the TDCIS wave function in the sense of the TD variational principle. The numerical experiments for the 1D helium, beryllium, and carbon atoms verified this theoretical fact. By analyzing the TD behavior of electron densities and orbitals, it was shown that single-orbital excitation scheme in the TD-RASSCF-S method is important to take into account the electron correlation especially around the nucleus during the interaction with laser fields. 
By the reduction of the number of orbitals and configurations, the TD-RASSCF-S method is obviously more applicable than the MCTDHF method for large systems.

\begin{acknowledgments}
It is a pleasure to thank Dr. Jeppe Olsen (Aarhus University) and  Dr. Lasse Kragh S{\o}rensen (Aarhus University) for useful discussions. This work was supported by the Danish Research Council (Grant No. 10-085430) and an ERC-StG (Project No. 277767-TDMET).
\end{acknowledgments}

\nocite{*}
\bibliography{aipsamp}


\end{document}